\documentstyle[prl,aps,graphics,epsfig]{revtex}

\input{psfig.sty}
\begin{document}
\pagestyle{empty}
\draft
\twocolumn
\wideabs{
\title{Fast, efficient error reconciliation for quantum cryptography}
\author{W. T. Buttler, S. K. Lamoreaux, J. R. Torgerson, G. H. 
Nickel, C. H. Donahue and C. G. Peterson}
\address{University of California, Los Alamos National Laboratory, Los Alamos, 
New Mexico 87545}
\date{\today}
\maketitle
\begin{abstract}{We describe a new error reconciliation protocol {\it Winnow} 
based on the exchange of parity and Hamming's ``syndrome'' for $N-$bit subunits
of a large data set. {\it Winnow} was developed in the context of quantum key 
distribution and offers significant advantages and net higher efficiency 
compared to other widely used protocols within the quantum cryptography 
community. A detailed mathematical analysis of Winnow is presented in the 
context of practical implementations of quantum key distribution; in 
particular, the information overhead required for secure implementation is one 
of the most important criteria in the evaluation of a particular error 
reconciliation protocol. The increase in efficiency for Winnow is due largely 
to the reduction in authenticated public communication required for its 
implementation.}

\end{abstract}
}
\narrowtext
\section{Introduction}

Quantum cryptography \cite{ref:qkd-protocols} presents special problems in 
regard to error correction of noisy quantum communications. Under the 
constraint that the public channel can be authenticated, and the assumption 
that all public communications can be eavesdropped, classical information on 
the exchanged qubits must be revealed through a series of public discussions 
to test the quantum key integrity and to remove the errors. Discrepancies 
within the qubits, observed as errors, must be treated as having been 
introduced by a hostile eavesdropper; the eavesdropper is generally referred 
to as Eve and labeled {\bf E} in this work. 

In a classical environment {\it all} errors can {\it always} be removed with 
the condition that to remove all errors one may have to reveal all 
information. However, within the secrecy framework imposed by quantum key 
distribution (QKD), revealed information reduces privacy and the effective 
channel capacity. Because of this great care must be taken to reveal a minimal 
amount of information to remove errors from quantum key while accounting for 
the leaked information to ensure key integrity after errors are removed.

Within this context of QKD, the two parties that exchange qubits over a 
quantum channel (Alice ({\bf A}) and Bob ({\bf B}) is the notation typically 
used within the quantum cryptography community) must have a fast and efficient 
method to mend the quantum key; in addition, they must also reduce {\bf E}'s 
knowledge gained during public discussions to a vanishingly small amount. 
These constraints require that any error reconciliation protocol will also 
need supporting protocols to provide a complete framework for quantum 
cryptographic security. That is, a useable QKD system will comprise a 
quantum-key transmitter ({\bf A}) and receiver ({\bf B}), and a series of 
protocols to remove errors and account for and mitigate the information 
leakage attributable to {\bf E}. The series of protocols includes 
\cite{ref:cm97,ref:lutkenhaus99}, but is not necessarily limited to the 
following: error-reconciliation \cite{ref:bbbss,ref:cascade}, privacy 
amplification \cite{ref:priv-amp} and signature authentication 
\cite{ref:sig-auth}.

In addition to these protocols, we acknowledge a protocol generally formulated 
in \cite{ref:bbbss} that we refer to as privacy maintenance. We also note that 
the predecessor to CASCADE \cite{ref:cascade} --- the best known and probably 
the most widely used error reconciliation protocol --- is also generally 
formulated in \cite{ref:bbbss} and is characterized by a binary search; here 
we refer to the binary search, which is a major element of CASCADE, as BINARY. 
A fundamental difference between BINARY and CASCADE is that CASCADE neglects 
privacy maintenance: all data are retained until the necessary privacy 
amplification is performed on the error-free data. We observe that the 
reconciliation process is more efficient if privacy maintenance is 
implemented during reconciliation as will become obvious in the following 
discussion.

Finally, this work introduces a new error reconciliation protocol that uses a 
Hamming code \cite{ref:Hamming1,ref:Hamming2} to remove errors. We refer to 
this protocol as {\it Winnow}. {\it Winnow} is characterized by the 
application of a parity test, a conditional Hamming hash, and privacy 
maintenance.

\section{Hamming Error Detection and Correction}

The application of the Hamming hash function for error correction 
\cite{ref:Hamming1,ref:Hamming2} is illustrated as follows: 

First, after {\bf A} and {\bf B} exchange qubits on the quantum channel, {\bf 
A} and {\bf B} then divide their random bits into blocks of length $N_h = 2^m 
- 1$. (Due to the 1:1 correlation of these data, we henceforth refer to these 
blocks as a single data- or bit-block.) The $m-$bit ($m \ge 3$) {\em 
syndromes} $S_a$ and $S_b$ are then calculated, where $S_a$ and $S_b$ 
respectively depend only on {\bf A}'s or {\bf B}'s bits in a particular block.

Next, {\bf B} transmits his syndrome to {\bf A} and errors are only discovered 
if the syndrome difference $S_d$ ({\em exclusive or} of $S_a$ with $S_b$) is 
non-zero:
\begin{equation}
  S_d = S_a \oplus S_b \neq \{0\}^m \text{.}
\label{eqn:syndrome-difference}
\end{equation}

Finally, $m$ bits are deleted from each bit block to eliminate the potential 
loss of privacy to {\bf E} due to the (classical) communication of {\bf B}'s 
syndromes: $m$ bits of information are revealed on each block for which $S_b$ 
is revealed reducing the channel capacity per symbol by 
$m/N_h$\cite{ref:shannon48}. 

Specifically, data privacy is maintained by removal of $m$ bits from each 
block at the $\{2^j\}$ positions where $j \in \{0 ,\ldots,m - 1 \}$. 
These bits are independent in the syndrome calculations as seen below in 
the matrix $h^{(m)}$:
\begin{equation}
h^{(3)} = \left[
\begin{array}{ccccccc}
1 & 0 & 1 & 0 & 1 & 0 & 1 \\ 
0 & 1 & 1 & 0 & 0 & 1 & 1 \\ 
0 & 0 & 0 & 1 & 1 & 1 & 1 \\
\end{array}
\right ] \text{,}
\label{eqn:h37}
\end{equation}
where for this particular matrix, $m \equiv 3$. We refer to the operation of 
discarding bits in this manner \cite{ref:bbbss} as {\it privacy maintenance}.

As a final comment on Eq. \ref{eqn:h37}, note that the transpose of 
$h^{(3)} \equiv \left [ h^{(3)} \right ]^T$ are the binary equivalent numbers 
$1$ to $7$, and is generalized such that 
$[h^{(m)})]^T \equiv \{1, \ldots, (2^m - 1)\}$, $N_h$ binary numbers. 

The matrix $h^{(m)}$ is a special form of hash function \cite{ref:GHN} and 
is represented by:
\begin{equation}
  h^{(m)}_{i,j} = \left \lfloor {j \over 2^{i-1}} \right \rfloor  
                  \left ( {\rm mod} \; 2 \right ) \text{,}
\label{eqn:h}
\end{equation}
where $i \in \{ 1,\ldots,m \}$, and $j \in \{ 1,\ldots,N_h \}$; arithmetic is 
{\it modulo 2}.

The Hamming algorithm always corrects any single error within any $N_h$-bit 
block, but the effect of the Hamming algorithm, which is related to the 
{\it syndromes} and privacy maintenance, is less clear in the event that 
more than one error exists in a bit block. Such considerations are now 
discussed in detail in terms of the syndromes.

The syndromes $S_a$ and $S_b$ are formed by contraction of the $N_h-$bit
blocks with the matrix $h^{(m)}$:
\begin{equation}
   S_i = \left ( \sum_{j=1}^{N_h} X_j h^{(m)}_{i,j} \, \right ) \left 
               ( {\rm mod} \; 2 \right) \in \{ 0,1 \}^m \text{,}
\label{eqn:syndromes}
\end{equation}
where subscript $i$ represents syndrome bit $i$ in the $m$-bit binary 
syndrome, $X_j$ represents bit {\it j} $\in$ {\bf A}'s or {\bf B}'s block, 
and $S = \{ S_i \}$ is the binary syndrome value of either {\bf B}'s or 
{\bf A}'s block. Understanding the effect of the syndromes in locating and 
correcting errors is crucial to assessing the performance of {\it Hamming}, 
and thus {\it Winnow}. 

The syndrome difference (Eq. \ref{eqn:syndrome-difference}) defines a binary 
number that gives the location of a single bit in {\bf A}'s or {\bf B}'s code 
word that when toggled from $0 \mapsto 1$ or from $1 \mapsto 0$ affects the 
syndrome difference $S_d$ such that when the syndrome difference is 
recalculated it gives the binary number $S_d^\prime \equiv \{0\}^m$. For 
example, if $S_d \ne \{ 0 \}^m$, then $S_d$ is an $m$-bit binary number whose 
value gives the location of a single bit in either {\bf A}'s or {\bf B}'s 
code-word to add {\it exclusive or} with the orignal bit value. After that 
bit value is changed, then the new syndrome for that code word is then 
calculated ({\it e.g.} $S_A^\prime$) and added (again, {\it exclusive or}) to 
the original syndrome for the other code word ($S_B$ in this example). The 
result is that the changing of the single bit indicated by the non-zero 
syndrome difference in the one code-word either corrects an error, or 
introduces another, in that code word. This is no great mystery but rather 
reflects the fact that {\it Hamming} codes are {\it n-k codes}. In this case, 
$n = 2^m - 1$ relates the number of bits in each code word ($N_h$), and 
$k = n - m$ relates the channel capacity (the channel capacity is 
$k/n \Longleftrightarrow k/N_h$ per bit) given the code (a {\it Hamming} code 
in this discussion). 

In an {\it n-k Hamming code}, there are $2^{(2^m)}$ unique code words 
characterized by $2^m$ unique syndromes; further, there are $2^k$ code words 
with the same syndrome. Because this code can correct $1$ error, it has a 
minimum {\it Hamming} distance of $d = 3$. This also means it can detect at 
least $2$ errors. In fact the {\it Hamming} distance $d$ for the {\it 
Hamming} code is $d \equiv 3$. 

By definition, a code word with a single error will have $S_d \ne \{ 0 \}^m$ 
(can obviously detect a single error if it can correct a single error). In 
addition, if a code word has exactly $2$ errors then by definition 
$S_d \ne \{ 0 \}^m$ (can detect at least $2$ errors if it can correct a 
single error). Therefore, if a code word has exactly $2$ errors, then after 
applying the {\it Hamming} algorithm, and after changing the bit value 
indicated by $S_d$, the code word will finish with exactly $3$ errors. The 
proof is by contradiction: If a code word with $2$ errors finished with 
$1$ error (an error was corrected), then the new syndrome difference would 
be non-zero! Contradiction also proves that $1$-error is corrected if there is 
exactly $1$ error: If an error was introduced the syndrome difference would 
again be non-zero. Thus, in examining {\it Hamming} codes we observe that a 
code word with $1$ error will finish with $0$ errors, but a code word with 
exactly $2$ errors finishes with exactly $3$ errors. In each case the new 
syndrome difference changes such that $S_d^\prime = \{ 0 \}^m$. 

By symmetry, if an $N_h$-bit code word contains exactly $N_h - 1$-errors (all 
the bits except one are in error), then after application of {\it Hamming} 
all the bits in the code word will be in error. Further, a code word that 
contains $N_h - 2$ errors will finish with $N_h - 3$ errors, {\it i.e} one of 
the errors is corrected. 

The above arguments imply that a {\it Hamming} code only works well if 
the probability of $2$ or more errors is low relative to the liklihood of a 
single, or no, errors. In either case the {\it Hamming} code is 
inefficient as $m$-bits are revealed in the syndrome (this fact is discussed 
in detail later).

The difficult question to answer in analyzing the performance of a 
{\it Hamming} code is how does {\it Hamming} affect code words with more than 
$2$, but less than $2^{m - 1}$, errors?

It is not obvious but the number of code words with $3$-errors and 
$S_d \equiv \{ 0 \}^m$ is related to the number of ways $2$-error code words 
map to a code word with $3$ errors (and $S_d = \{ 0\}^m$). In other words, 
there must be a way to arrange $3$ errors in a code word and still maintain 
$S_d = \{ 0 \}^m$. Lacking this would mean that the code could always detect 
more than $2$ errors with a {\it Hamming} distance of $d = 3$.

To complete the {\it Hamming} efficiency analysis, how code words with $3$ 
or more errors are affected after application of {\it Hamming} must be 
analyzed. For $3$ errors it is now obvious: there must be at least $2^m - 1$ 
ways to start with $3$ errors in an $N_h$-bit code word and still finish with 
$3$ errors. In the case that there exist $3$ errors in a code word, and 
$S_d \ne \{ 0 \}^m$, then an error will be introduced into the $N_h$-bit code 
word because if the code word finished with $2$ errors then 
$S_d \ne \{ 0 \}^m$---a contradiction.

As a special case (example), consider $m = 3$. There are 
${7 \choose 3} = 35$ ways to arrange $3$ errors in $7$ bits. Because there are 
exactly $7$ non-zero syndrome differences for $m = 3$ and $n_i = 2$, there 
must be {\it at least} $7$ ways to arrange $3$ errors in $7$-bits and have 
$S_d \equiv \{ 0 \}^m$. In fact for this special case this is the result. 
What this means is that, statistically, $7$ in $35$ code words with $3$-errors 
will finish with $3$-errors, and $28$ in $35$ words with $3$ errors will 
finish with $4$ errors. Thus, code words that start with $3$ errors will 
finish with $19/5$ errors per $7$-bit block, in the limit of an infinite 
number of $7$-bit blocks with exactly $3$ errors. By symmetry, it is obvious 
that given an infinite number of $7$-bit blocks with exactly $4$ errors, the 
final error rate per block would be $16/5$---a lower final error rate.

Thus, what is needed is a way to calculate, for any number $m$ of parity 
checks, in {\it Hamming}, a way to calculate the number of ways to arrange 
the initial number of errors per block and finish with $S_d = \{ 0 \}^m$, or 
with $S_d \ne \{ 0 \}^m$. Eq. \ref{eqn:prob} permits that calculation for 
any initial number of errors per block, $n_i$, given any initial block size, 
$N_h$:
\begin{eqnarray}
%\label{eqn:prob}
   N_{S_d \ne 0} + N_{S_d = 0} & = & {N_h \choose n_i} \nonumber \\
  -N_{S_d \ne 0} + N_h \cdot  N_{S_d = 0} & = & (-1)^q \cdot N_h \cdot 
                                 {\frac{N_h - 1}{2} \choose p} \\
\label{eqn:prob}
   \Longleftrightarrow 
                {\left [ \matrix{N_{S_d \ne 0} \cr N_{S_d = 0}} \right ]} 
         & = & \left [ \matrix{ N_h & 1 \cr 
                                        -1 & 1 \cr } \right ]^{-1} 
                       \left [ \matrix {{N_h \choose n_i} \cr
                                              (-1)^q {\frac{N_h - 1}{2}
                                                  \choose p} \cr }\right ]
                    \nonumber  \text{,}
%\label{eqn:prob}
\end{eqnarray}
where $q = \lceil n_i/2 \rceil$, $p = \lfloor n_i/2 \rfloor$, $n_i$ is the 
initial number of errors per {\it Hamming} block of $N_h = 2^m - 1$ bits per 
block; in this situation, $N_{S_d = 0}$ gives the number of syndrome 
differences with $S_d = \{ 0 \}^m$, and $N_{S_d \ne 0}$ gives the number of 
syndrome differences with $S_d \ne \{ 0 \}^m$. Eq. \ref{eqn:prob} is 
generalized by dividing both sides by the total number of ways to arrange 
$n_i$ errors in the $N_h$ bits. In this situation we find a more useful 
quantity:
\begin{eqnarray}
   \Pi_{S_d = 0} & = & \frac{N_{S_d = 0}}{{N_h \choose n_i}} \text{, and} 
                                                              \\ \nonumber
   \Pi_{S_d \ne 0} & = & \frac{N_{S_d \ne 0}}{{N_h \choose n_i}} \text{.}
\label{eqn:Pi_Sd0}
\end{eqnarray}
This result is required later. 

These arguments are not obviously general for the case of $m > 3$, but they 
give insight into the general problem. The difficulty with the special case of 
$m = 3$ and $n_i = 3$ is that the next case of $n_i = 4$ is symmetric and 
complementary with $n_i = 3$, as mentioned previously. Further, as was noted, 
there is no path to map $3$ errors to $2$ errors as $S_d \ne \{ 0 \}^m$ when 
there are exactly $2$ errors. However, Eq. \ref{eqn:prob} is the general 
technique to calculate the quantities specified, {\it i.e.} the number of 
ways to map $n_i$ errors to $S_d = \{ 0 \}^m$ or not, given $N_h = 2^m - 1$ 
bits in a block.

Given these facts, how the errors change for 
$m \ge 4$ and $4 \le n_i < 2^{(m - 1)}$ is the general result of interest. 

Let $n_i^{(+)}$ be the number of ways to increase the number of errors from 
$n_i$ to $n_i + 1$, in a bit-block, and $n_i^{(-)}$ the number of ways to 
decrease the number of errors from $n_i$ to $n_i - 1$; of course, the 
considerations relate to $m \ge 4$. The results are as follows: 
\begin{eqnarray}
   n_i^{(+)} & = & N_{S_d = 0}(N_h | n_i) + (n_i + 1) \cdot 
                            N_{S_d = 0}(N_h | n_i + 1) \nonumber \\
   n_i^{(-)} & = & {N_h \choose n_i} - n_i^{(+)} \text{,} 
\end{eqnarray}
where $N_{S_d = 0}(N_h | n_i + 1)$ is the number of ways to arrange $n_i + 1$ 
errors in $N_h$ bits and obtain $S_d = \{ 0 \}^m$ (the reader will recall that 
earlier it was stated that the number of ways to get $S_d = \{ 0 \}^m$ for 
$n_i + 1$ errors is directly related to the number of ways to map 
$n_i \mapsto n_i + 1$ errors); of course, $N_{S_d = 0}(N_h | n_i)$ is the 
number of ways to arrange $n_i$ errors in $N_h$ bits and get 
$S_d = \{ 0 \}^m$. Thus, the generalized probability for the number of errors 
$n_i$ to increase, or decrease is:
\begin{eqnarray}
   \Pi^{(+)} & = & \frac{n_i^{(+)}}{n_i^{(+)} + n_i^{(-)}} \nonumber 
                                                            \text{, and} \\
   \Pi^{(-)} & = & 1 - \Pi^{(+)} \text{.}
\label{eqn:Pi_pm}
\end{eqnarray}

\section{Winnow}

As a general rule, the ideal error correcting protocol would correct all bit 
errors in each bit block, introduce no additional bit errors, and reveal a 
minimal amount of information on the key bits to an eavesdropper through 
public communication. The outlined {\it Hamming} protocol has a number of 
shortcomings regarding this ideal. First, the difference syndrome $S_d$ does 
not distinguish between single- and multiple-bit errors. Therefore, additional 
errors may be introduced if instances of $S_d \neq \{ 0 \}^m$ are treated as 
due to single errors. Second, up to $m$ bits of information are exchanged for 
each data block reducing channel capacity per symbol with each exchange: 
information which can be compromised by eavesdropping.

One solution is to eliminate all bits within data blocks for which 
$S_d\neq\{0\}^m$. This certainly removes the possibility of introducing 
additional bit errors into the key, but, unfortunately, the efficiency of such 
a method is low as {\it every} block loses either $m$-bits to privacy
maintenance, or all bits because $S_d \neq \{0\}^m$. The efficiency of this 
approach is not optimal as most of the discarded bits/blocks for which $S_d 
\neq \{ 0 \}^m$ are probably not in error.

Another, more powerful solution is to introduce a preliminary parity 
comparison on a block of $N = 2^m$ bits and to make a comparison of the 
syndromes $S_a$ and $S_b$ conditional upon the result of the parity 
comparison.\footnote{Hamming discusses the addition of a parity check on the
$N_h = 2^m - 1$ bit block \cite{ref:Hamming2} (pp. 47-48; pp. 213-214). His 
conclusion is that {\bf A} and {\bf B} are more likely to introduce additional 
errors than correct errors by changing a bit if $S_d \neq \{ 0 \}^m$ and the 
block-parities agree. In this situation {\bf A} and {\bf B} could either 
remove the $m + 1$ bits required to ensure privacy on the remaining bits 
(which may remove errors), or they could eliminate all of the bits in 
question, as $n_i \in \{ 2, 4, \ldots, 2^m - 2 \} > 1$. The expanded protocol 
described in this effort allows the detection of an even or odd number of 
errors and prevents a correction attempt on those data blocks with even 
numbers of errors. This is important since the Hamming algorithm will 
increase the number of errors in blocks which have $2 \le n_i \le N_h/2$.}

If the block parities do not agree an odd number of errors exists in the 
$N$-bit block. Moreover, if the bit errors are distributed randomly 
throughout the data, and if the number of errors is sufficiently small, then 
an odd number of errors in a block probably indicates a single error which 
can be corrected by the additional application of the Hamming algorithm. For 
example, in the situation that a block contains {\it one} bit error, if 
$S_d = \{ 0 \}^m$ then the first bit is in error. (By symmetry it is clear 
that if there are exactly $N - 1$ errors in the block the first bit would not 
be in error.) Thus, this approach always allows the correction of a single 
error in the $N$ bits, {\it i.e.} if the bits are to be retained. However, in 
the protocol outlined here the one bit is regularly discarded for privacy 
maintenance (for the exchanged parity bit) and the Hamming algorithm is 
applied to the remaining $N_h$ bits, as previously discussed, and then 
$\lceil \log_2(N_h) \rceil$ additional bits are discarded to complete the 
privacy maintenance giving a channel capacity of $(2^m - m - 1)/N$ per symbol 
on blocks that contain an initial parity error. This appears to be an 
additional loss of channel capacity, but because the syndromes are not 
exchanged and compared when the block parities agree the channel capacity 
actually increases over the basic Hamming algorithm; one bit is still 
discarded from the blocks that do not exhibit a parity error for privacy 
maintenance. We refer to this error reconciliation protocol as {\it Winnow}. 

{\it Winnow} reveals $\log_2(N) + 1$ bits in $2$ classical communications 
when the parities on the $N$ bits do not agree: $m$ bits for the 
syndrome and $1$ bit for parity; conversely, {\it Winnow} reveals $1$ bit of 
information in $1$ classical communication when the parities 
agree.\footnote{Exchanging the parity on $N = 2^m$ bits instead of 
$N_h = 2^m - 1$ bits results in slightly higher channel capacity. That is: 
more information is revealed when the syndrome information is combined with 
the parity information on a $N_h$ bit block than is revealed when the parity 
and syndrome are revealed on $N$ bits in {\it Winnow}.}

Therefore, the amount of key data discarded is
\begin{equation}
  N^{odd}_{dis.} = \log_2(N) + 1 = m + 1
\end{equation}
bits for blocks with odd numbers of errors such that the fraction of the bits 
remaining after privacy maintenance is
\begin{equation}
  \mu^{odd}_{pm} = 1 - \frac{N^{odd}_{dis.}}{N} \text{.}
\end{equation}
For $N \in \{ 8, 16, 32, 64, 128 \}$, 
$\mu^{odd}_{pm} \in \{ 0.5, 0.69, 0.88, 0.89, 0.94\}$, respectively. Also,
\begin{equation}
  \mu^{even}_{pm} = 1 - \frac{1}{N} \text{,}
\end{equation}
and $\mu^{even}_{pm} \in \{ 0.88, 0.94, 0.97, 0.98, 0.99 \}$ for the same 
values of $N$. In either case, the appropriate overhead for the classical 
communications is also removed immediately from the data so that the privacy 
of the bits is at least maintained if not improved.

All single bit errors in an $N$-block are guaranteed to be either eliminated 
or corrected after a single pass of {\it Winnow} (a {\it Winnowing}). What 
remains to be considered is how blocks with multiple errors affect the 
overall efficiency of {\it Winnow}.

\section{Winnow Efficiency}

Define the change in number of errors in a given block and {\it for a given 
initial number of errors} as $\Delta n = n_f - n_i$, where $n_i$ and $n_f 
\equiv n_f(n_i|N)$ are the initial and final numbers of bit errors in a block 
prior to and after {\it Winnowing}, respectively. The average change in the 
number of errors, for a given number of initial errors, after a 
{\it Winnowing} (this step includes elimination of the parity bit but not the 
final $m$-bits required for completion of the privacy maintenance step) can 
be expressed as
\begin{equation}
  \bar{\Delta n} \equiv \big\langle \! \Delta n(n_i) \! \big\rangle = 
                       \sum_{\Delta n= -2}^{1} \Delta
       n \! \cdot \! p(\Delta n | n_i) \text{,}
\label{eqn:edn}
\end{equation}
where
\begin{equation}
    \sum_{\Delta n = -2}^{1}p(\Delta n | n_i) = 1 \text{,}
\end{equation}
and $p(\Delta n|n_i)$ is the probability that the number of errors will change 
by $\Delta n \in \{-2, -1, 0, 1\}$ given an initial condition of $n_i$ errors 
in an $N-$bit data block. The $p(\Delta n|n_i)$ of interest can be written 
more instructively as
\begin{eqnarray}
  p(\! +1|n_i) \! & = & \! \pi^{(n)} \! \! \cdot \! \Pi_{S_d \neq 0}(n_i) \! 
         \cdot \! \Pi^{(\! + \!)}(n_i) \nonumber \\
  p(\! \pm 0 |n_i) \! & = & \! \pi^{(n)} \! \! \cdot \! \Pi_{S_d = 0}(n_i) \! 
         + \! \pi^{(y)} \! \cdot \! \Pi_{S_d \neq 0}(\delta n_i) \! \cdot \! 
             \Pi^{(\! + \!)}(\delta n_i) \nonumber \\
  p(\! -1|n_i) \! & = & \! \pi^{(n)} \! \! \cdot \! \Pi_{S_d \neq 0}(n_i)
          \! \cdot \! \Pi^{(\! - \!)}(n_i) + \pi^{(y)} \! \cdot \! \Pi_{S_d = 
             0}(\delta n_i) \nonumber \\
  p(\! -2|n_i) \! & = & \! \pi^{(y)} \! \!\cdot\!\Pi_{S_d \neq 0}(\delta n_i)
          \! \cdot \! \Pi^{(\! - \!)}(\delta n_i) \text{,}
\end{eqnarray}
where, $n_i$ is as previously defined, $\delta n_i \equiv n_i -1$, $\pi^{(y 
\vee n)}$ depends only on the initial number of errors ($n_i$) in the $N$-bit 
block and is the probability the bit discarded for privacy maintenance 
following the parity check was ({\it y}), or ($^\vee$) was not ({\it n}) in 
error; $\Pi_{S_d = 0}(n_i ^\vee \delta n_i)$ and $\Pi_{S_d \neq 0}(n_i ^\vee 
\delta n_i)$ are the probabilities that $S_a \! = \! S_b$ or $S_a \! \neq \! 
S_b$ for  $n_i$ or $\delta n_i$ errors in $N_h$ bits and are concretely 
defined in Eq. \ref{eqn:Pi_Sd0}, and $\Pi^{(\pm)}(n_i ^\vee \delta n_i)$ 
is defined in Eq. \ref{eqn:Pi_pm}.

Eq. \ref{eqn:edn} can be expressed in terms of $\pi^{(y \vee n)}$, 
$\Pi_{S_d}$ and $\Pi^{(\pm)}$ as
\begin{eqnarray}
    \bar{\Delta n} & \equiv &
     \big\langle \! \Delta n^{(n)}(n_i) \! \big\rangle
    + \big\langle \! \Delta n^{(y)}(n_i) \! \big\rangle \text{,}\nonumber \\
    & = & \bar{\Delta n}^{(n)} + \bar{\Delta n}^{(y)} \text{,}
\label{eqn:fdn1}
\end{eqnarray}
where the arguments which depend on $n_i$ have been suppressed, and
\begin{eqnarray}
  \bar{\Delta n}^{(n)} \! \! & = & \pi^{(n)} \! \! \cdot \! \Pi_{S_d \neq 
        0}(n_i) \! \cdot \! \Big[1 \! - \! 2 \! \cdot \! \Pi^{(-)}(n_i)\Big] 
        \text{,} \nonumber \\
  \bar{\Delta n}^{(y)} \! \! & = & \pi^{(y)} \! \cdot \! \Pi_{S_d \neq 
        0}(\delta n_i) \! \cdot \! \Big[1 \! - \! 2 \! \cdot \! 
        \Pi^{(-)}(\delta n_i) \Big]  \! - \pi^{(y)} \text{.}
\label{eqn:fdn2}
\end{eqnarray}

The final quantity needed to calculate the efficiency of {\it Winnow} is 
$\pi^{(y \vee n)}$:
\begin{eqnarray}
   \pi^{(y)} & = & \frac{n_i}{N} \nonumber \text{, where} \\
   \pi^{(y)} + \pi^{(n)} & = & 1 \nonumber \text{.}
\label{eqn:eq}
\end{eqnarray}

Table \ref{tab:nf} and Table \ref{tab:pf} provide a concrete example for the 
special case of $m = 3$ of the effects of {\it Winnow} on blocks with exactly 
$n_i \in \{0, \ldots, 8\}$ errors. Table \ref{tab:nf}, 
introduces a new quantity
\begin{equation}
  \bar{n}_f \equiv \langle n_f \rangle = n_i + \bar{\Delta n} \text{,}
\end{equation}
and in Table \ref{tab:pf} a new parameter
\begin{equation}
    p_f = \frac{\bar n_f}{N_f} 
\end{equation}
is defined. 

The parameter $p_f$ defines the probability for each bit in a given block to 
be in error. The number $N_f \in \{N - 1, N - m - 1\}$ and its value depends 
on the action required by {\it Winnow} for a given number of initial errors. 
For example, $N_f = N - 1$ or $N - m - 1$ for $p_f$ and $n_i$ even or odd, 
respectively.

These two tables illustrate the effect of {\it Winnow} on data which are 
divided into 8-bit blocks. The values marked with superscript $p$ reflect the 
effect of discarding one bit following the parity comparison. The values 
marked with superscript $ph$ refer to the data after the Hamming algorithm is 
also applied, but before the requisite $\log_2(N) = 3$ bits of data are 
discarded for privacy maintenance. The final values denoted by subscript $f$ 
reveal the effect of {\it Winnow} (including the effect of all discarded data 
required for privacy maintenance).

The parameter $p_f$ clearly shows a reduction in errors for $n_i = 1$ and an 
increase in errors for $n_i = 3$. It also shows that discarding data to 
maintain privacy of the remaining key has no effect on the error probability. 

%\subsection{Probability for Residual Errors}

The fraction of key remaining after a {\it Winnowing} is given by
\begin{equation}
    \mu_N \equiv \frac{\langle N_f \rangle}{N}
            =    \frac{\sum_{n_i = 0}^N \, N_f \, P(n_i|N)}{N} \text{,}
\label{eqn:mue}
\end{equation}
and the probability for any key bit to be in error following a {\it Winnowing} 
is
\begin{eqnarray}
  p_N = \frac{ \langle \bar{n}_f \rangle}{\langle 
                           N_f \rangle}
      = \frac{ \sum_{n_i = 0}^N \bar n_f(n_i) \! \cdot \! P(n_i|N)}
                                            {N \! \cdot \! \mu_N} \text,
\label{eqn:pe}
\end{eqnarray}
where $P(n_i|N)$ is the probability for an $N$-bit block to contain $n_i$ 
errors before a {\it Winnowing}.

%\subsubsection{Random Distribution of Errors}

Obviously, the efficiency with which {\it Winnow} removes errors depends upon 
the distribution of errors within the data. Without intimate knowledge of a 
specific QKD apparatus, a reasonable assumption is that the errors are random 
and normally distributed throughout the data. Given this assumption, 
$P(n_i|N)$ in Eq. \ref{eqn:pe} is given by the binomial distribution
\begin{equation}
  P(n_i\mid N,p_0) = {N \choose n_i}{p_0}^{n_i}(1 - p_0)^{N - n_i}
\end{equation}
where $p_0$ is the probability that any given bit is in (relative) error.

With this assumption, Eqs. \ref{eqn:mue} and \ref{eqn:pe} can be expressed as
\begin{eqnarray}
  \mu_N = \frac{N - 1 - m \sum_{n_i^{odd}}{N \choose n_i}{p_0}^{n_i}(1 - 
                                                   p_0)^{N - n_i}}{N} \text,
\label{eqn:muebi}
\end{eqnarray}
where $m = \log_2(N)$, and 
\begin{eqnarray}
   p_N = \frac{\sum_{n_i = 0}^N \bar n_f(n_i){N\choose n_i}{p_0}^{n_i}(1 - 
                                                            p_0)^{N - n_i}}
  {N \! \cdot \! \mu_N} \text{.}
\label{eqn:pebi}
\end{eqnarray}
%

%\section{Analysis}

The efficiency with which {\it Winnow} reduces errors in the key is of great 
interest. Two related issues which concern the efficiency are: 1) the number 
of iterations of {\it Winnow} necessary to achieve a sufficiently low 
probability of error in the remaining key data, and 2) the amount of key data 
that is discarded through privacy maintenance. 

The number of iterations is of concern because each iteration reveals 
information and consumes time with each communication between {\bf A} and {\bf 
B}. Moreover, each communication requires the use of some private key for 
signature authentication \cite{ref:sig-auth}. Most importantly, though, is 
that each iteration requires a significant amount of data to be discarded 
through privacy maintenance.

\begin{figure}
\centerline{\psfig{figure=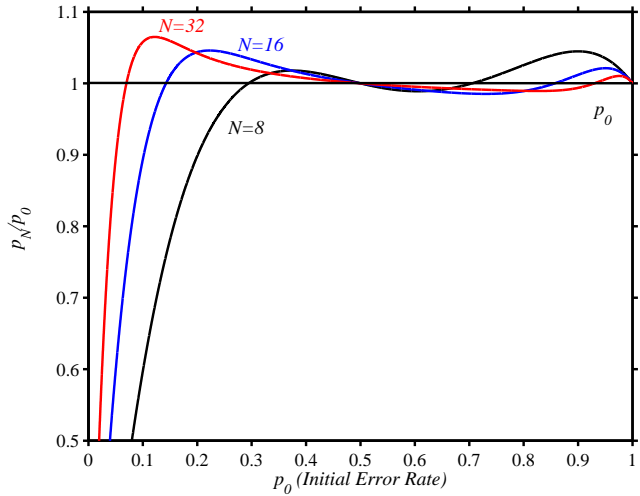,width=3.375 in}}
\caption{The ratio $p_N/p_0$ for $N = 8$, $16$ and $32$. These curves 
illustrate the change in  the probability that a given bit is in error after a 
single application of {\it Winnow} for the indicated block size $N$. Note that 
$( p_8 < p_{16} ) \; {\forall} \; (p_0 < 0.38)$; in addition, 
$(p_{16} < p_{32}) \; {\forall} \; (p_0 < 0.20)$. This indicates that a
pplications of {\it Winnow} with smaller $N$ are more efficient at removing 
errors than are applications with larger $N$ within the region where $p_0$ 
satisfies these conditions.}
\label{fig:justin}
\end{figure}

Smaller $N$ require more data to be discarded than larger $N$ as can be seen 
from Eq. \ref{eqn:muebi}. However, an effect which tends to mollify this 
undesirable condition is that smaller $N$ are more efficient at removing 
errors for larger values of initial error probability. This effect is 
illustrated in Fig. \ref{fig:justin} where we have plotted $p_N/p_0$ for 
several values of $N$. For all values of $N$ and $p_0$ sufficiently small, 
$p_N/p_0 < 1$ and the protocol can remove errors from the key data. However, 
as $p_0$ increases from $p_0 = 0$, each of the curves passes through 
$p_N/p_0 = 1$ indicating that additional errors are being introduced into the 
key. Moreover, the value of $p_0$ for which $p_N/p_0 = 1$ is smaller for 
larger N and the curves do not intersect between $p_0 = 0$ and $p_N/p_0 = 1$.

As a primary requirement of {\it Winnowing} real data in an iterative 
application, a random shuffling of the data between iterations is essential to 
randomly redistribute missed or introduced errors. Without this random shuffle 
multiple errors remain clumped together and, in essence, are impossible to 
completely remove from the data. Under this constraint it is obvious that the 
final error probability, and the amount of data remaining after a number of 
{\it Winnowings}, depends on the way in which $N$ is varied throughout the 
successive {\it Winnowings}. An intuitive result which we have verified 
empirically is that less data are discarded for the same initial and final 
error probabilities if $N$ is chosen well for the first iteration and is 
either held constant or increased for all subsequent iterations; there is no 
advantage to decreasing $N$ in subsequent iterations if {\it Winnow} is 
applied as outlined here. 

Define
\begin{equation}
  p(p_0; \{ j_N \})
\end{equation}
and
\begin{equation}
  \mu(p_0; \{ j_N \})
\end{equation}
as the final error rate and fraction of data remaining after a sequence 
$\{j_N\} = \{j_8,j_{16},j_{32},j_{64},j_{128}\}$ where $j_N$ iterations of 
{\it Winnow} are applied with a block size $N \in \{8,16,32,64,128\}$ 
beginning with $N = 8$ and increasing monotonically in $N$ by factors of 
$2$.\footnote{In this work $N$ is constrained such that $N \le 128$ only for 
the sake of brevity. We have found that this constraint does not impose a 
serious limit on the ability of {\it Winnow} to correct errors. The ideas 
discussed below can be extended to include $N > 128$ in a straightforward 
manner.}

Because $(p_8 < p_0) \; {\forall} \; (p_0 < 0.5)$, it may appear that errors 
can be corrected in the data for this entire range of initial error 
probability. However, there is another criterion that must be met which 
significantly reduces the maximum correctable error probability: There must 
remain a finite amount of error-free data after the potential information 
possessed by {\bf E} is reduced through privacy amplification.

The maximum amount of potential information possessed by {\bf E} can be 
determined by the initial error probability $p_0$ and depends on the QKD 
protocol and the type of attacks being employed. For example, if the BB84 
protocol is used and {\bf E} employs a complete intercept/resend attack on the 
quantum channel in the same bases used by {\bf B}, she will introduce an error 
probability of $p_0=1/4$. She will also potentially know $1/2$ of the data 
before error reconciliation and up to $2/3$ of the data which remains after 
error reconciliation.

If {\bf E} uses a more clever intercept/resend strategy of detecting and 
resending in the Breidbart basis (second paper in \cite{ref:bbbss}), she would 
introduce the same number of errors ($p_0=1/4$) and could know up to a 
fraction of $0.59$ of the data before error reconciliation and $0.78$ of the 
data remaining after error reconciliation.

It should also be noted that certain states of light are more susceptible to 
attack than others. For example, consider weak coherent states which are 
commonly used in QKD systems. If {\bf E} also employs a beamsplitter attack 
\cite{ref:lutkenhaus99,ref:bbbss,ref:dhh99} against one of these systems, an 
additional amount of data is compromised which is not greater than the mean 
number of photons in the state. However, this value can be made arbitrarily 
small so it is neglected in the following calculations. Moreover, other states 
of light can be used in QKD schemes which are not vulnerable to this type of 
attack \cite{ref:BSAttackPaper}.

Thus, the fraction of data remaining after error reconciliation and
privacy amplifications can be 
\begin{equation}
  \nu^{bb84} = \mu - (0.59) 4 \, p_0
\label{eqn:nubb84}
\end{equation}
for BB84, where $\nu$ describes the remaining fraction of key.

From the above considerations, $p$ and $\nu$ can be investigated as a function 
of $p_0$. Of particular interest is the maximum $p_0$ for which some secure 
data remains while achieving a sufficiently low final error probability to 
make the data useful. We have chosen, somewhat arbitrarily, $p \le10^{-6}$ as 
a reasonable target for the final error probability.

With this target and the remaining fraction of private data described by Eq. 
\ref{eqn:nubb84}, we find the largest initial error probability for which some 
private data remains is 
\begin{equation}
    p_0 = 0.1322 \text{,}
\end{equation}
after {\it Winnowing} and privacy amplification.

To achieve $p \lesssim 10^{-6}$ from this large initial error probability, 
{\it Winnow} must be applied in the sequence $\{j_N\} = \{3, 1, 0, 1, 3\}$. 
That is, $3$ {\it Winnowings} with $N = 8$ must be followed by $1$ 
{\it Winnowing} with $N = 16$, {\it etc.} If this prescription is followed, 
\begin{equation}
  \nu^{bb84} = 0.0017
\end{equation}
of the original data remain and are secure following privacy amplification.

Some QKD schemes require a larger estimate of {\bf E}'s knowledge. If Eq. 
\ref{eqn:nubb84} is replaced with \cite{ref:bbbss}
\begin{equation}
  \nu = \mu - 2 \sqrt{2} \, p_0 \text{,}
\end{equation}
we find
\begin{equation}
  p_0 = 0.1222
\end{equation}
for  $\{j_N\} = \{3, 0, 1, 0, 4\}$. This leaves a fraction $\nu = 0.0017$ of 
the original data as secure data with a single-bit error probability 
$\le 10^{-6}$.

Finally, if we estimate that {\bf E} knows every bit of data by causing 
$p_0 = 1/4$, then
\begin{equation}
  \nu = \mu - 4 \, p_0 \text{.}
\end{equation}
We then find that the largest reconcilable $p_0$ is
\begin{equation}
  p_0 = 0.1037
\end{equation}
for $\{ j_N \} = \{2, 1, 1, 0, 3\}$ and $\nu = 0.0020$.

The most efficient iteration sequence ($\{ j_N \}$) for any QKD scheme can be 
determined by first applying {\it Winnow} with $N = 8$ to estimate $p_0$. Once 
the number of blocks with odd and even (even includes zero) errors, 
$M^{odd}_e$ and $M^{even}_e$ respectively, are known, the fraction 
\begin{equation}
   \frac{\# \text{ of Parity Errors}}{\# \text{ of Blocks}} 
    = \frac{{\sum_{n_i^{odd}}{N\choose n_i}{p_0}^{n_i}(1 - p_0)^{N - n_i}}}{N}
\end{equation}
can be used to estimate $p_0$. Knowledge of $p_0$ is sufficient to determine 
the $\{j_N\}$ which maximizes $\nu$. 

For small $p_0$, the most efficient $\{j_N\}$ may start with $N > 8$. However, 
working systems that have been reported in the literature 
\cite{ref:bbbss,ref:working-qkd} have large enough error probabilities so that 
the most key is left if $N = 8$ for at least the first iteration.

A detailed analysis of the advantages of {\it Winnow} over other protocols is 
beyond the scope of this work. However, it is instructive to note the 
advantages over at least the best-known protocol CASCADE. 

The most notable dif\-fer\-ence be\-tween {\it Winnow} or BINARY and CASCADE 
is that CASCADE does not employ privacy maintenance. The disadvantage of such 
a protocol is that super-redundant information must be exchanged with each 
successive iteration. This is to be compared with BINARY and {\it Winnow} 
which reduce the size of the data set with each communication. With the 
reasonable requirement that a bit revealed through these communications 
requires at least a bit to be eliminated through some channel, either before 
or during privacy amplification, then the inefficiency of keeping all bits 
until all errors are removed becomes obvious: retaining and repetitively 
exchanging information on the same bits is an additional expense to the 
protocol. 

For the purpose of comparison, we have computed the maximum $p_0$ which BINARY 
(less privacy maintenance) can successfully reconcile errors and preserve a 
small amount of secure data after privacy amplification and the removal of the 
super-redundant information. We find
\begin{equation}
   p_0 = 0.114
\end{equation}
for $\{j_N\} = \{2, 1, 0, 2, 1\}$ and $\nu^{bb84} = 0.01$ when 
$(0.59) 4 \, p_0$ describes the additional amount of key that must be 
discarded through privacy amplification. This is to be compared with 
$p_0 = 0.1322$ for the same considerations with {\it Winnow}. This 
application of BINARY is a reasonable approximation to CASCADE which may 
include a higher order correction giving a slightly higher overall error 
reduction than BINARY without privacy maintenance.

This comparison (or any of the previous discussion) does 
not take into account bits used to authenticate messages sent between {\bf A} 
and {\bf B}. Both CASCADE and BINARY requires significantly more two-way 
communication than {\it Winnow}, and each packet of $n$ bits sent may require 
$\lceil \log_2(n) \rceil$ for authentication \cite{ref:sig-auth}. We 
calculate that the most efficient application of CASCADE requires a minimum 
of  $1 + \log_2(N)$ communications per iteration while {\it Winnow} requires 
only $2$ communications {\it for any block size} $N$ that exhibits a parity 
error; the additional communications required imposes a tight limitation on 
practical efficiency. In addition, because CASCADE does not maintain privacy, 
subsequent iterations requires more bits to be exchanged in the initial 
parity phase with each iteration. The additional bit exchanges may require 
additional signature authentication bits. 

We acknowledge that because CASCADE and BINARY always removes a single 
error and never introduces additional errors to multiple error blocks, both 
BINARY and CASCADE perform infinitesimally better than {\it Winnow} in an 
environment where signature authentication is not required and privacy 
maintenance is removed from the {\it Winnow} and BINARY protocols. However, 
{\it Winnow's} $2$ communications is a great advantage where time is of 
the essence with regard to production of secure key bits over inefficient 
noisy quantum channels. 

\section{Conclusion}

We have identified a new, fast, efficient, error reconciliation protocol for 
quantum key distribution which requires only $2$ communications between the 
two parties attempting to reconcile private, quantum key material. We refer to 
this protocol as {\it Winnow}.

{\it Winnow} incorporates a preliminary parity comparison on blocks whose 
size is $N = 2^m$ where $m \in \{3, 4, 5, 6, ...\}$. Subsequently, one bit is 
discarded from these blocks to maintain the privacy of the remaining bits. A 
{\it Hamming} hash function, which can be used to correct single errors, is 
applied to the remaining $N - 1$ bits on the blocks whose parities did not 
agree. Finally, $m$ bits are discarded from the blocks on which the Hamming 
algorithm was applied to maintain the privacy of those bits.

We find this protocol capable of correcting an initial error probability of up 
to $13.22\%$ in privacy amplified BB84-like quantum key distribution schemes. 

\vspace{5mm}
{\noindent {\bf Acknowledgments:} The authors extend their thanks and 
appreciation to R. J. Hughes, E. Twyffort, D. P. Simpson and J. S. Reeve for 
many helpful discussions regarding this effort.}

\newpage
\begin{table} %[t]
\caption{$\bar{n}_f$ for $N = 8$ for various stages in {\it Winnow} (note that 
{\it Hamming} is not applied to blocks that contain an even number of errors).}
\begin{tabular}{|c|ccccccccc|}
  $n_i$
    & 0 & 1 & 2 & 3 & 4 & 5 & 6 & 7 & 8 \\ \hline
  $\bar{n}_f^p$
    & 0 & 0.88 & 1.75 & 2.63 & 3.5 & 4.38 & 5.25 & 6.13 & 7 \\ \hline
  $\bar{n}_f^{ph}$
    & 0 & 0 & 1.75 & 3.5 & 3.5 & 3.5 & 5.25 & 7 & 7 \\ \hline
  $\bar{n}_f$
    & 0 & 0 & 1.75 & 2.0 & 3.5 & 2.0 & 5.25 & 4 & 7 \\
\end{tabular}
\label{tab:nf}
\end{table}
\begin{table} [H]
\caption{$\bar n_f/N_f$ for $N = 8$ for various stages in {\it Winnow} (note 
that the Hamming component of {\it Winnow} is not applied to blocks that 
contain an even number of errors).}
\begin{tabular}{|c|ccccccccc|}
  $p_i$
    & 0 & 0.13 & 0.25 & 0.38 & 0.5 & 0.63 & 0.75 & 0.88 & 1 \\ \hline
  $p_f^p$
    & 0 & 0.13 & 0.25 & 0.38 & 0.5 & 0.63 & 0.75 & 0.88 & 1 \\ \hline
  $p_f^{ph}$
    & 0 & 0 & 0.25 & 0.5 & 0.5 & 0.5 & 0.75 & 1 & 1 \\ \hline
  $p_f$
    & 0 & 0 & 0.25 & 0.5 & 0.5 & 0.5 & 0.75 & 1 & 1 \\
\end{tabular}
\label{tab:pf}
\end{table}


\begin{thebibliography}{99}
%
\bibitem{ref:qkd-protocols} C. H. Bennett and G. Brassard, {\it International 
Conference on Computers, Systems \& Signal Processing, Bangalore, India, 1984} 
(IEEE, New York, 1984) 175-179; 
A. K. Ekert, Phys. Rev. Lett. {\bf 67}, 661-663 (1991); 
C. H. Bennett and S. J. Wiesner, Phys. Rev. Lett. {\bf 69}, 2881-2884 (1992); 
C. H. Bennett, Phys. Rev. Lett. {\bf 68}, 3121-3124 (1992).
%
\bibitem{ref:cm97} C. Cachin and U. M. Maurer, J. Cryptology {\bf 10}, 97-110 
(1997).
%
\bibitem{ref:lutkenhaus99} N. L$\ddot{\mathrm{u}}$tkenhaus, Phys. Rev. A 
{\bf 59}, 3301-3319 (1999).
%
\bibitem{ref:bbbss} C. H. Bennett {\it et al.}, Lect. Notes in Comput. Sci. 
{\bf 473}, 253-265 (1990); J. of Cryptology {\bf 5}, 3-28 (1992).
%
\bibitem{ref:cascade} G. Brassard and L. Salvail, Lect. Notes Comput. Sci. 
{\bf 765}, 410-423 (1994).
%
\bibitem{ref:priv-amp} C. H. Bennett {\it et al.}, IEEE Trans. Inf. Theory 
{\bf 41}, 1915-1923 (1995).
%
\bibitem{ref:sig-auth} W. Diffe and M. E. Hellman, Proceedings of AFIPS 
National Computer Conference, 109-112 (1976); 
R. L. Rivest, A. Shamir and L. M. Adleman, Communications of the ACM {\bf 21}, 
120-126 (1978); 
C. Mitchell, F. Piper and P. Wild, G. J. Simmons (Ed.), {\it Contemporary 
Cryptography: The Science of Information Integrity}, 325-378 IEEE Press, 1992.
%
\bibitem{ref:Hamming1} R. W. Hamming, The Bell System Technical Journal 
{\bf 2}, 147-161 (1950).
%
\bibitem{ref:Hamming2} R. W. Hamming, {\it Coding and Information Theory}, 
Prentice Hall, 239 pp, New Jersey (1986,1980).
%
%\bibitem{ref:shannon48} C. E. Shannon, ``A mathematical theory of 
%communication,'' The Bell System Technical Journal {\bf 27}, 379-423 and 
%623-656 (1948).
%
\bibitem{ref:shannon48} C. E. Shannon, The Bell System Technical Journal 
{\bf 27}, 379-423 and 623-656 (1948).
%
\bibitem{ref:GHN} A. J. Menezes, P. C. van Oorschot and S. A. Vanstone, {\it 
Handbook of Applied Cryptography}, 780 pp, CRC Press, New York (1997).
%
\bibitem{ref:dhh99} M. Du$\check{\mathrm{s}}$ek, O. Haderka and M. Hendrych, 
Opt. Commun. {\bf 169}, 103-108 (1999).
%
\bibitem{ref:BSAttackPaper} J. Kim {\it et al.}, Nature {\bf 397} 500-503 
(1999); 
P. Michler {\it et al.}, Science {\bf 290}, 2282 (2000); 
B. Lounis and W. E. Moerner, Nature {\bf 407}, 491 (2000).
%
\bibitem{ref:working-qkd} J. D. Franson and H. Ives, Appl. Opt. {\bf 33}, 
2949-2954 (1994); 
B. Jacobs and J. D. Franson, Opt. Lett. {\bf 21}, 1854-1856 (1996); 
P. D. Townsend, Nature {\bf 385}, 47-49 (1997); 
A. Muller, H. Zbinden and N. Gisin, Europhys. Lett. {\bf 33}, 335-339 (1996); 
W. T. Buttler {\it et al.}, Phys. Rev. Lett. {\bf 84}, 5652-5655 (2000); 
R. J. Hughes, G. L. Morgan and C. Glen Peterson, J. Mod. Opt. {\bf 47}, 
533-547 (2000); 
D. S. Bethune and W. P. Risk, IEEE J. Quant. Elect. {\bf 36}, 340-347 (2000); 
M. Bourennane {\it et al.}, J. Mod. Opt. {\bf 47}, 563-579 (2000); 
J. G. Rarity, P. R. Tapster and P. M. Gorman, J. Mod. Opt. {\bf 48}, 
1887-1901 (2001).
%
\end{thebibliography}
\end{document}